\documentclass[letter]{aa} 
\usepackage{amsmath}
\usepackage{graphicx}

\usepackage{txfonts}

\usepackage{color}

\begin{document}

\title{Model selection with the Pantheon+ Type Ia SN sample}

\author{N. Chandak$^1$, F. Melia$^2$\thanks{John Woodruff Simpson Fellow} and
  J. Wei$^{3}$}

\offprints{N. Chandak}
\titlerunning{The Pantheon+ Type Ia Sn sample}
\authorrunning{Chandak, Melia \& Wei}

\institute{$^1$Department of Physics, The University of Arizona, Tucson, Arizona 85721, USA\\
$^2$Department of Physics, The Applied Math Program, and Department of Astronomy,
The University of Arizona, Tucson, Arizona 85721, USA; \email{fmelia@email.arizona.edu} \\
$^3$Purple Mountain Observatory, Chinese Academy of Sciences, Nanjing 210023, China
}
   \date{Received November 25, 2025}

  \abstract
   {Recent discoveries, e.g., by JWST and DESI, have elevated the level of
   tension with inflationary $\Lambda$CDM. For example, the empirical evidence
   now suggests that the standard model violates at least one of the energy
   conditions from general relativity, which were designed to ensure that
   systems have positive energy, attractive gravity and non-superluminal energy flows.}
   {In this Letter, we use a recently compiled Type Ia supernova sample to
   examine whether $\Lambda$CDM violates the energy conditions in the local Universe,
   and carry out model selection with its principal competitor, the $R_{\rm h}=ct$
   universe.}
   {We derive model-independent constraints on the distance modulus based on
   the energy conditions and compare these with the Hubble diagram predicted
   by both $\Lambda$CDM and $R_{\rm h}=ct$, using the Pantheon+ Type Ia supernova catalog.}
   {We find that $\Lambda$CDM violates the strong energy condition over the redshift
   range $z \subset (0, 2)$, whereas $R_{\rm h}=ct$ satisfies all four energy
   constraints. At the same time, $R_{\rm h}=ct$ is favored by these data
   over $\Lambda$CDM with a likelihood of $\sim 89.5\%$ versus $\sim 10.5\%$.}
   {The $R_{\rm h}=ct$ model without inflation is strongly favored by the Type Ia
   supernova data over the currrent standard model, while simultaneously adhering 
   to the general relativistic energy conditions at both high and low redshifts.}


   \maketitle

\section{Introduction}\label{introduction}

Inflationary $\Lambda$CDM
\citep{1979JETPL..30..682S,1981PhRvD..23..347G,1982PhLB..108..389L}
has been quite successful in accounting for many cosmological observations over the past
few decades, most notably the multi-peak structure in the CMB angular power spectrum.
But though $\Lambda$CDM, like essentially all cosmological models today, is based
on the Friedmann-{Lema$\hat{\rm i}$tre}-Robertson-Walker (FLRW) metric, it is
largely an empirical model, relying on eleven to twelve observationally optimized
parameters. New data acquired with Planck \citep{2020A&A...641A...6P},
DESI \citep{2024AJ....168...58D} and, most recently, JWST
\citep{2022ApJ...936L..14P, 2022ApJ...940L..55F, 2024A&A...690A...2M, 2024Natur.633..318C},
have raised major concerns regarding its status as a `true' standard model
\citep{Melia:2020,Melia:2025}. For example, the discovery of supermassive black holes
\citep{2024PDU....4601587M} and well-formed galaxies \citep{2023MNRAS.521L..85M}
merely a few hundred million years after the Big Bang refute its predicted timeline,
affirming other inconsistencies and growing tension at lower redshifts.

A well-known inconsistency in this model is the discordant measurement of the
Hubble Constant at high and low redshifts. Planck observes a value
$67.4 \pm 0.5 \: \rm km\,s^{-1}\,Mpc^{-1}$ \citep{2020A&A...641A...6P},
which disagrees by more than $7\sigma$ with the value determined using
Cepheid-calibrated Type Ia SNe, i.e., $73.04 \pm 1.04 \: \rm km\,s^{-1}\,Mpc^{-1}$
\citep{2022ApJ...934L...7R}. The uncertainties in these measurements have dropped
as the precision has improved, so it is unlikely that the discrepancy is due to
systematics. It appears that some unknown physics beyond $\Lambda$CDM is responsible
for this tension.

In contrast, the $R_{\rm h}=ct$ cosmology avoids virtually all of the inconsistencies
and flaws in $\Lambda$CDM \citep{Melia:2020,Melia:2025}. Over 35 different
comparative tests have now been completed, showing that $R_{\rm h} = ct$ is
favored over $\Lambda$CDM at all redshifts. This alternative FLRW model is better
motivated theoretically, and adheres to all the known constraints from
general relativity, thermodynamics and particle physics. For example, it
alone satisfies the zero active mass condition, which is required for a
proper use of the FLRW ansatz.

An additional strong factor in its favor is that it does not require inflation
to explain the uniformity of the CMB temperature across the sky
\citep{2013A&A...553A..76M}, while still fully adhering to the energy conditions.
On the other hand, an inflationary spurt from $t=10^{-35} \: \rm s$ to
$10^{-32} \: \rm s$ is required in $\Lambda$CDM to solve the horizon problem,
though the inflaton field violates at least one of the energy conditions
in the early Universe.

The purpose of this Letter is to demonstrate that the early Universe is
not the only time when $\Lambda$CDM violates these important energy constraints
in general relativity. Type Ia SNe, HII~Galaxies, and cosmic chronometers
have all hinted that $\Lambda$CDM violates them at low redshifts as well
\citep{2007PhRvD..75h3523S}, while $R_{\rm h}=ct$ does not
\citep{Chandak:2025}. Here, we use the latest Type Ia SN catalog for model
selection between these two cosmologies, and robustly test their fits against the
energy conditions at $z\lesssim 2$.

\section{Background}\label{background}

Type Ia SNe have been used previously for model selection between
$\Lambda$CDM and $R_{\rm h}=ct$ \citep{2015AJ....149..102W, 2018EL....12359002M}.
But we now have access to the best spectroscopically confirmed measurements in the Pantheon+ Sample \citep{2022ApJ...938..113S},
which will be used in our analysis. The novelty in our approach, however,
is that we also introduce model-independent bounds on the distance modulus
based on the strong energy condition (SEC) from general relativity. The
energy conditions were introduced to ensure that the chosen stress-energy
tensor in Einstein's equations does not contain negative energy densities 
(the Null and the Weak Energy Conditions), and to ensure 
that the source of gravity is attractive (the Strong Energy 
Condition) and does not lead to superluminal energy flows 
(the Dominant Energy Condition).

The $R_{\rm h}=ct$ universe \citep{2007MNRAS.382.1917M, 2012MNRAS.419.2579M}
has emerged as the most likely competitor to the current standard model. 
It has by now been tested with 35 different data sets, 
successfully accounting for these observations better than $\Lambda$CDM, 
often with likelihoods $\sim 90-95\%$ versus $\sim 5-10\%$ (a recent 
comprehensive list of tests may be found in \citealt{Melia:2020} and
\citealt{Melia:2025}). Moreover, not only does $R_{\rm h} = ct$ account 
for the data better, it does so while adhering to all known fundamental 
physical constraints, including the zero active mass condition required 
for the validity of the FLRW ansatz \citep{2022MPLA...3750016M}.
One of its principal advantages is that it requires only one free 
parameter---the Hubble constant, $H_0$---while $\Lambda$CDM relies on 
the optimization of $11-12$ different variables.

The energy conditions from general relativity have most famously been 
applied to the singularity theorems of Penrose \citep{Penrose:1965} and 
Hawking \citep{HawkingEllis:1973} and to the positive mass theorem 
\citep{Schoen:1979}. As we shall see, the energy condition most relevant 
to the present analysis is the Strong Energy Condition (SEC),
which is formally derived as follows.

The SEC requires gravity to be attractive, i.e that matter 
gravitates towards matter. The inflaton field, e.g., easily violates this
condition because it produces antigravity. When invoking the inflationary
concept, one must acknowledge the fact that, unlike every other known
particle and field, it and a cosmological constant would be the only
physical entities generating antigravity. The SEC can be represented 
mathematically via a convergence constraint on the Ricci Tensor, 
$R_{\mu \nu}$, for a timelike observer with four velocity 
$u^{\mu}$ \citep{Martin-Moruno:2017},
\begin{equation}\label{SEC_definition}
R_{\mu \nu} \: u^{\mu} u^{\nu} \leq 0.
\end{equation}
Using Einstein's field equations, this can be converted into a constraint 
on the stress-energy tensor itself,
\begin{equation}
\bigg ( T_{\mu \nu} - \frac{1}{2}Tg_{\mu \nu} \bigg ) \:u^{\mu} u^{\nu} \geq 0,
\end{equation}
where $g_{\mu \nu}$ are the metric coefficients and $T=T_{\mu \nu}\,g^{\mu \nu}$ 
is the trace of $T_{\mu\nu}$. Using the perfect fluid approximation, 
with $T_{\mu \nu} = {\rm diag} \:(\rho, p, p, p)$), this condition leads 
to the following constraint(s) on the total pressure $p$ and energy density 
$\rho$ of the system \citep{2023AnP...53500157M}:
\begin{equation} \label{SECequiv}
\rho + p\geq 0, \:\:\:\:\:
\rho + 3p\geq 0\,.
\end{equation}

In an FLRW universe, the SEC constrains the expansion to be
non-accelerating. Hence, any acceleration in the standard model is a direct 
violation of this energy condition. One sometimes encounters the argument 
that such a violation by inflation thus invalidates the SEC. But assigning 
priority to inflation over this GR constraint appears to be premature, given 
that the reality of inflation has never been established conclusively. In fact, 
modern high-precision measurements tend to disfavor inflation (see, e.g., 
\citealt{2013PhLB..723..261I,Liu:2020}). 

The fact that the SEC is violated both by an inflaton field and dark energy in 
the guise of a cosmological constant means that $\Lambda$CDM is inconsistent 
with classical general relativity at both high and low redshifts. For 
a more detailed explanation about the application of energy conditions to 
cosmology, see \cite{2023AnP...53500157M}, \cite{Chandak:2025} and the
comprehensive account in \cite{Melia:2025}.

In this Letter, we shall demonstrate the seriousness of this violation 
based on Equations~(1) and (3) at $z\lesssim 2$. But dark energy need not be
a cosmological constant. In the $R_{\rm h}=ct$ model, it is dynamic, presumably
an extension to the standard model of particle physics. This is why the cosmic
expansion at all redshifts is fully consistent with the energy conditions in
$R_{\rm h}=ct$, as opposed to the tension created in the standard model.

\section{Data and analysis}\label{data}

The Pantheon+ Sample contains 1701 spectroscopically confirmed
Type Ia SNe, compiled from over 18 different surveys (see Table 1 of
\citealt{2022ApJ...938..113S}). The data release was accompanied by 
a description of the empirical light-curve fits that are used to determine the peak
magnitudes $x_0$, light-curve shapes $x_1$, colors $c$, and the time of peak
brightness $t_0$, of each SN. These can be used with a background
cosmology to generate distance moduli using the modified
Tripp relation \citep{1998A&A...331..815T},
\begin{equation}\label{distance_modulus_observed}
        \mu_{\rm obs} = -2.5\log_{10}(x_0) + \alpha x_1 - \beta c + M_B\,,
\end{equation}
where $\alpha$, $\beta$, and $M_B$ are 3 model-dependent nuisance parameters to be
optimized simultaneously with the background cosmology.

\begin{figure}[t]
        \centerline{
                \includegraphics[angle=0,scale=0.34]{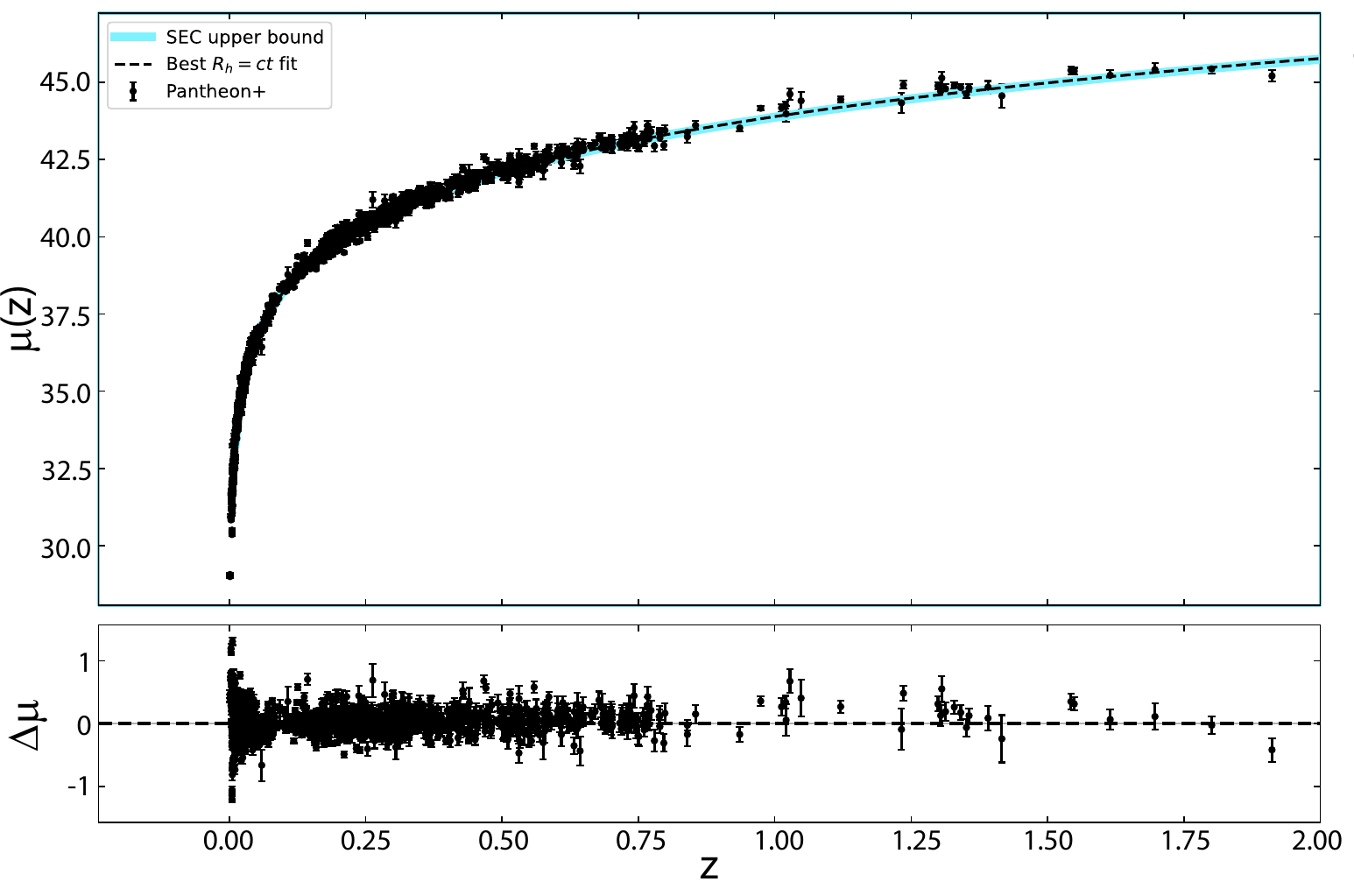}}
        \caption{Distance modulus for the Type Ia SNe in the $R_{\rm h}=ct$ model.
	The predicted (dashed) curve is coincident with the SEC limit (blue curve),
        as highlighted in the plot (Figure~\ref{f4}) without the data.}
        \label{f1}
\end{figure}

\begin{figure}[h]
        \centerline{
                \includegraphics[angle=0,scale=0.31,trim={0 0 0 0},clip]{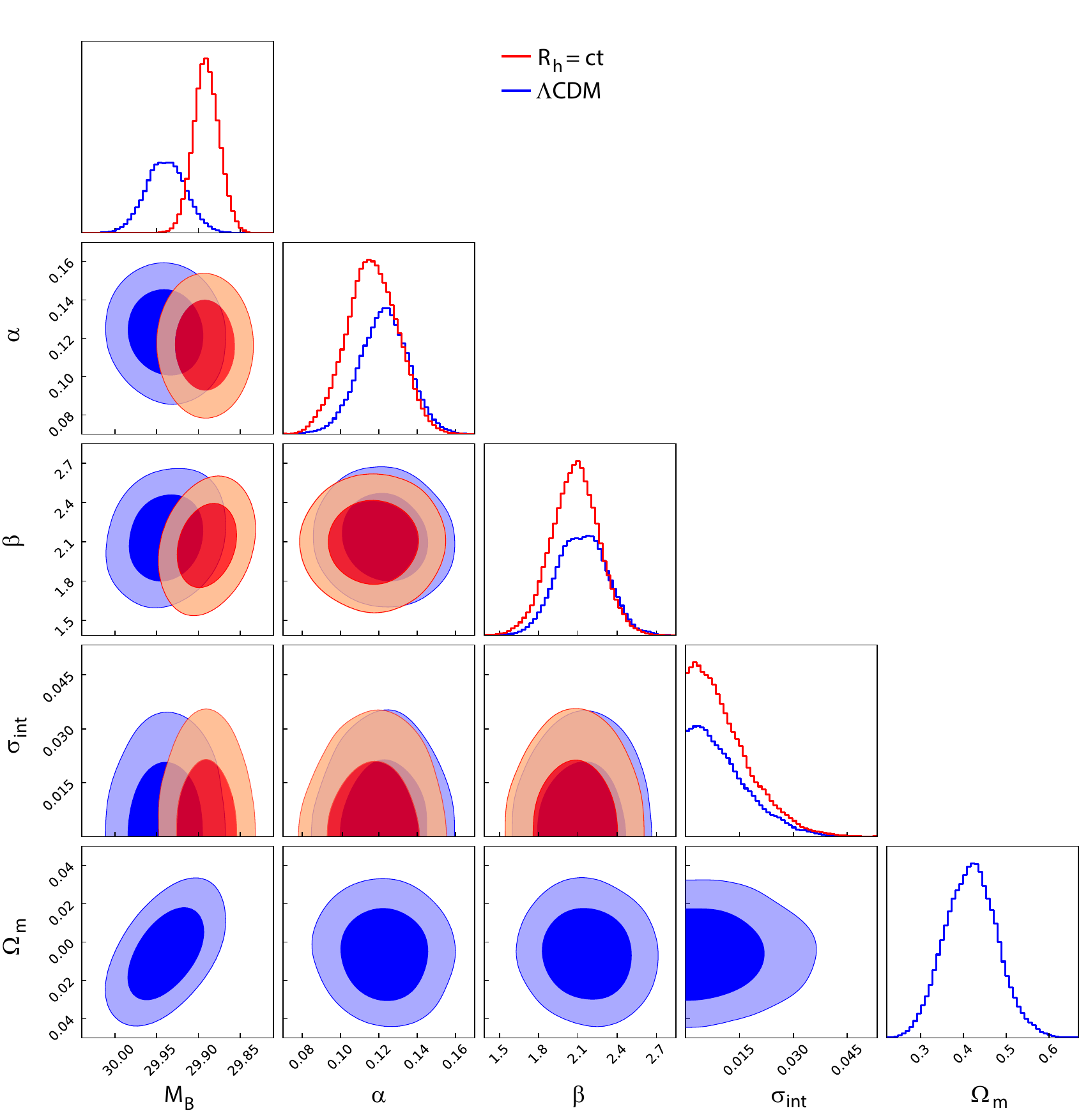}}
        \caption{1D probability distributions and 2D regions with the $1 - 2\sigma$
        contours for the parameters $M_B$, $\alpha$, and $\beta$ in $\Lambda$CDM (blue)
        and in $R_{\rm h}=ct$ (red), corresponding to the fits in Figures~\ref{f1} 
	and~\ref{f3}. Note that not all best fit values in both models are the same.}
        \label{f2}
\end{figure}

For each SN, the theoretical distance modulus is
\begin{equation}\label{distance_modulus_theoretical}
\mu_{\rm th}(z) = 5\log\bigg[\frac{D_L (z)}{Mpc} \bigg] + 25\,,
\end{equation}
where $D_L(z)$ is the model-dependent luminosity distance in terms of the
redshift $z$. In flat-$\Lambda$CDM,
\begin{equation}\label{LCDM_luminosity_distance}
D_L^{\Lambda {\rm CDM}}(z) = \frac{c}{H_0} (1+z) \int_0^{z}{\frac{dz'}
{\sqrt{\Omega_m(1+z')^3 + (1-\Omega_m)}}}\,,
\end{equation}
assuming spatial flatness and a negligible radiation pressure.
In $R_{\rm h}=ct$, we have instead
\begin{equation}\label{Rh_luminosity_distance}
D_L^{R_h=ct}(z) = \frac{c}{H_0} (1+z) \ln(1+z)\,.
\end{equation}

We follow the procedure described in \cite{2015AJ....149..102W}, based on the method
of maximum likelihood estimation (MLE), to optimize the nuisance parameters
simultaneously with each cosmological model. However,
$H_0$ is simply set to $70$ km s$^{-1}$ Mpc$^{-1}$ in both cases
because its value is degenerate with $M_B$. In other words, these two
parameters may not be optimized separately because they are directly coupled via
Equations~(4) and (6). Our optimized value for $M_B$ is thus consistent with this 
fiducial choice.

We maximize the likelihood function
\begin{equation}\label{MLE}
\mathcal{L} = \frac{1}{\sqrt{(2\pi)^N \det C}} \times \exp\bigg [{-\frac{1}{2} \Gamma^T C^{-1} \Gamma}\bigg],
\end{equation}
where $\Gamma = (\mu_{\rm obs} - \mu_{\rm th})$, and $C$ is the full $(N \times N)$ (with $N$ the number of SNe) covariance matrix given by,
\begin{equation} \label{covariance_matrix}
C=C_{\rm stat} + C_{\rm sys}.
\end{equation}
Here, $C_{\rm stat}$ is the total statistical uncertainty recomputed for each cosmology
and $C_{\rm sys}$ is the covariance from systematic uncertainties provided as part of the
data release. The statistical uncertainties are computed including contributions from (i) the
uncertainties and covariances of the light curve fit parameters, $\sigma_{l.c}$; (ii) the
redshift uncertainties which were propagated onto the distance modulus,
$\sigma_{D_L(z_i)}$; (iii) the uncertainties in peculiar velocities, $\sigma_{\rm pec} =
5\sigma_{z, i}/(z_i \log 10)$; (iv) lensing effects, $\sigma_{\rm lens} = 0.055 \times z_i$
and (v) an intrinsic dispersion, $\sigma_{\rm int}$ that we optimize along with the nuisance
parameters. As it turns out, however, its value has no significant effect on our results. Out
of these, the first 2 are cosmology dependent, hence $C_{\rm stat}$ is computed
simultaneously with the optimization:
\begin{equation} \label{C_stat}
C_{\rm stat, i} = \sigma^2_{l.c., i} + \sigma^2_{D_L(z_i)} + \sigma^2_{\rm pec, i} + \sigma^2_{\rm lens, i} + \sigma^2_{\rm int}
\end{equation}

\begin{figure}[t]
        \centerline{
                \includegraphics[angle=0,scale=0.34]{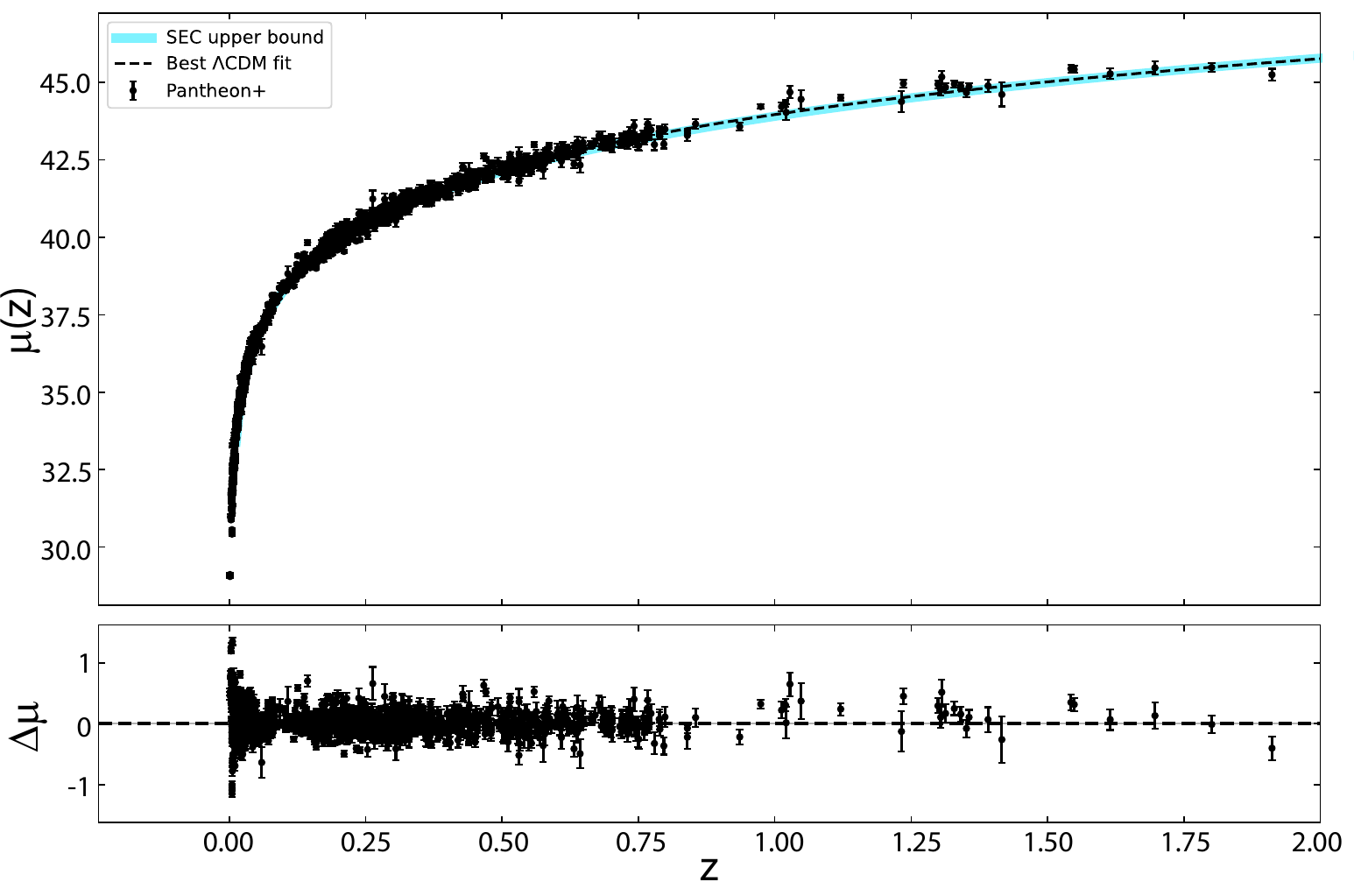}}
        \caption{Distance modulus for the Type Ia SNe in flat-$\Lambda$CDM.
        The model prediction (dashed curve) violates the SEC limit (blue curve)
        in the redshift range $z \subset (0.125, 2)$, as shown more clearly without
        the data in Figure~\ref{f4}.}
        \label{f3}
\end{figure}

\begin{table*}
        \centering \caption{Optimized parameters and model selection}
        \begin{tabular}{lccccccc}
                \hline
                \hline
                &&&&&&&\\
                Model&$\alpha$&$\beta$&$\Omega_{\rm m}$&$M_B$&$\sigma_{\rm int}$&BIC&Probability  \\
                &&&&&&&\\
                \hline
                &&&&&&&\\
                $R_{\rm h}=ct$&$0.117\pm0.013$&$2.088\pm0.179$&$-$&$-29.890\pm0.015$&$0.011 \pm 0.008$
                &$1439.27$&$89.80\%$\\
                $\Lambda$CDM&$0.121\pm0.014$&$2.124\pm0.185$&$0.416\pm0.059$&$-29.940\pm0.023$&$0.011 \pm 0.008$
                &$1443.62$&$10.20\%$\\
                &&&&&&&\\
                \hline
        \end{tabular}
        \label{table1}
\end{table*}

\begin{figure}[t]
        \centerline{
        \includegraphics[angle=0,scale=0.62]{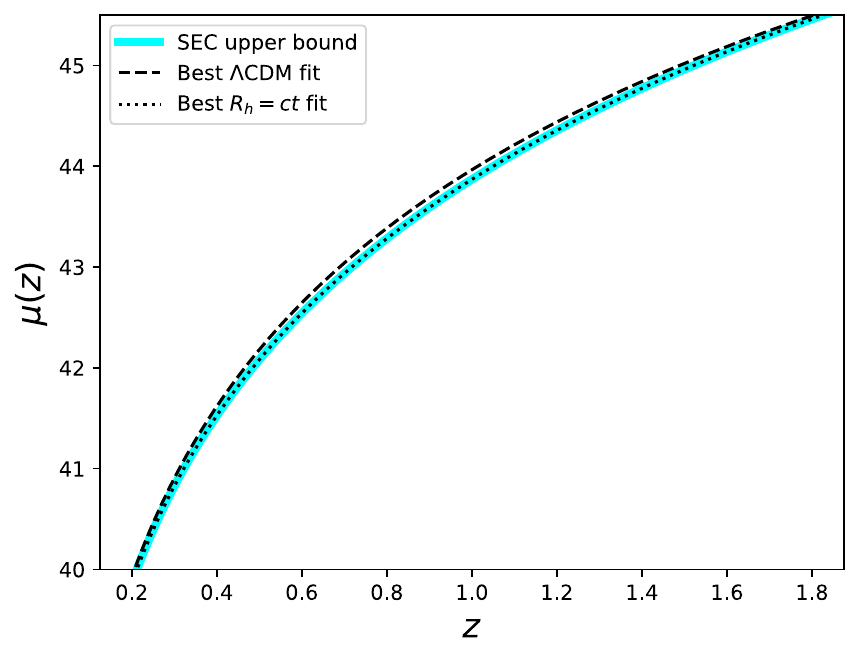}}
        \vskip -0.2in
        \caption{The optimized theoretical curves from Figures~\ref{f1}
        and \ref{f3}, together with the SEC upper bound (blue),
        showing full compliance by $R_{\rm h}=ct$ (dotted) and
        a violation by flat-$\Lambda$CDM (dashed).}
        \label{f4}
\end{figure}

We use the Python Markov chain Monte
Carlo (MCMC) module, EMCEE \citep{2013PASP..125..306F}, to generate the best fits in each model.
The best-fit Hubble diagrams and their residuals are shown in Figures~\ref{f1} and 
\ref{f3}, along with the upper bound to $\mu_{\rm th}$ derived from the SEC. These 
bounds on the distance modulus due to the energy conditions were derived in {\cite{Chandak:2025}}
The optimized parameters are shown in Table~\ref{table1}, and the corner plots are presented 
in Figure~\ref{f2}. The optimized fits and the SEC bound are repeated without the data 
and residual for greater clarity in Figure~\ref{f4}.

\section{Hubble diagrams and model selection}\label{Hubble diagrams and model selection}

An inspection of figures~(\ref{f1}) and (ref{fig3}), and a comparison of the error 
distributions in figure~(\ref{f2}), demonstrate that both models fit the data quite well. 
For example, the residuals in figures~(\ref{f1}) and (\ref{f3}) are virtually indistinguishable,
keeping in mind that the data are somewhat model dependent, since they rely in part on
the optimization of Equation~(4). In other words, the data themselves are not identical
in these two plots, but the residuals are very similar. The quality of these fits is 
also attested to by the optimized parameters shown in Table~1, and a side-by-side 
assessment of the error distributions in figure~(\ref{f2}). Not only are the (common) 
parameters very similar in the two models, but their $1-\sigma$ and $2-\sigma$ 
errors imply a comparable quality of the fits.

However, while both $R_{\rm h}=ct$ and flat-$\Lambda$CDM satisfy the null, weak 
and dominant energy conditions (not shown here, but see \citealt{Chandak:2025} for
more details) only $R_{\rm h}=ct$ also satisfies the SEC. The standard model 
clearly does not. In $\Lambda$CDM, dark energy is typically represented as 
a cosmological constant, which is known to violate the SEC.

But these fits reveal much more than merely this difference with respect to
the SEC. They show that the linear fit in $R_{\rm h}=ct$ is strongly favored 
by the Type Ia SN data over flat-$\Lambda$CDM, as indicated by the Bayes 
Information Criterion (BIC) \citep{1978AnSta...6..461S}. The $\Delta$BIC 
$\approx 4$ yields a probability of $\sim 89.8 \%$ for $R_{\rm h}=ct$ versus 
only $\sim 10.2 \%$ for flat-$\Lambda$CDM. The accelerated expansion predicted 
by $\Lambda$CDM is thus not favored by Type Ia SNe. This outcome
adds considerable weight to the already large body of evidence, drawn from 
much of the available data, that $R_{\rm h}=ct$ is favoured by the observations
over the current standard model. These include measurements based on HII Galaxies, 
cosmic chronometers, Compact Quasar Cores, the anglular diameter distance to lensed 
sources, the cosmic microwave background and other such sources.

\section{Conclusion}\label{conclusion}
Type Ia SNe have been key in uncovering the expansion of the local Universe and 
the existence of dark energy. They were also used to highlight the presumed 
acceleration at $z\lesssim 0.7$ in the context of flat-$\Lambda$CDM
\citep{1998ApJ...509...74G, 1998Natur.391...51P, 1998AJ....116.1009R, 1998ApJ...507...46S}.
But our updated analysis demonstrates that---while dark energy unquestionably
exists, though not as a cosmological constant---these data actually favor the 
linear expansion expected in $R_{\rm h}=ct$.

We have also affirmed the conclusion drawn in \cite{Chandak:2025},
based on the use of HII~Galaxies and cosmic chronometers, that an
accelerated expansion in the local Universe violates the SEC, while
the predicted Hubble diagram in $R_{\rm h}=ct$ is fully compliant
with all the energy conditions from general relativity.

The most compelling statement we can make is that an expansion adhering to 
the energy conditions actually provides the better fit to the Type Ia SN data, 
strengthening both the theoretical basis for the energy conditions themselves 
and the interpretation of the cosmological observations. Indeed, it is the model 
relying on speculative antigravity (i.e., $\Lambda$CDM) that is disfavored by 
the data.

\begin{acknowledgements}
FM is grateful to Amherst College for its support through a John Woodruff
Simpson Lectureship. This work is partially supported by the National Natural 
Science Foundation of China (grant Nos. 12422307 and 12373053).
\end{acknowledgements}

\bibliographystyle{aa}
\bibliography{a58990-26}

\label{LastPage}
\end{document}